\documentclass[letterpaper, 10pt]{ieeeconf} 
\IEEEoverridecommandlockouts
\usepackage{graphicx, xcolor, cite}
\definecolor{NCSUred}{RGB}{153, 0, 0}
\definecolor{NCSUgreen}{RGB}{0, 132, 115}
\definecolor{NCSUblue}{RGB}{65, 86, 161}
\definecolor{NCSUorange}{RGB}{209, 73, 5}
\usepackage[utf8]{inputenc}
\usepackage{setspace} 
\usepackage{amsmath, amssymb}
\usepackage{algorithm, algpseudocode}
\usepackage{hyperref}
\hypersetup{colorlinks=true, breaklinks=true, plainpages=false, citecolor=NCSUblue, linkcolor=NCSUgreen, urlcolor=NCSUblue}
\newtheorem{theorem}{Theorem}
\newtheorem{definition}{Definition}
\newtheorem{remark}{Remark}
\newtheorem{assumption}{Assumption}

\title{\LARGE \bf Data-Driven Observer Synthesis for Autonomous Limit Cycle Systems through Estimation of Koopman Eigenfunctions}
\author{Angela Ni\textsuperscript{1} and Wentao Tang\textsuperscript{1} 
\thanks{*This work is supported by NSF CBET \#2414369 and the Research Experience for Undergraduates grant by the College of Engineering, NC State University.}  
\thanks{$^{1}$ The authors are with the Department of Chemical and Biomolecular Engineering, North Carolina State University, Raleigh, North Carolina, U.S.A. Corresponding author: W. Tang
{\tt\small wentao\_tang@ncsu.edu}} 
}

\begin{document}
\maketitle
\thispagestyle{empty}
\pagestyle{empty}
\begin{abstract}
The signal of system states needed for feedback controllers is estimated by state observers. One state observer design is the Kazantzis-Kravaris/Luenberger (KKL) observer, a generalization of the Luenberger observer for linear systems. The main challenge in applying the KKL design is constructing an injective mapping of the states, which requires solving PDEs based on a first-principles model. This paper proposes a data-driven, Koopman operator-based method for the construction of KKL observers for planar limit cycle systems. Specifically, for such systems, the KKL injective mapping is guaranteed to be a linear combination of Koopman eigenfunctions. Hence, the determination of such an injection is reduced to a least-squares regression problem, and the inverse of the injective mapping is then approximated using kernel ridge regression. The entire synthesis procedure uses solely convex optimization. We apply the proposed approach to the Brusselator system, demonstrating accurate estimations of the system states. 
\end{abstract}

\section{INTRODUCTION}
State-feedback controllers, requiring system states as inputs, encounter the obstacle that some states are unmeasurable due to budget and technological constraints. This challenge motivates the construction of state observers to estimate unknown system states based on measured outputs. 
Typically, the observer is designed as a dynamical system, where the \emph{observer states} can be mapped to an asymptotically converging estimation of the true states \cite{Review, Review2}. 
The classical framework for asymptotically observing linear time-invariant systems is the \emph{Luenberger observer} \cite{Luen}, which is a copy of the system dynamics with an additional observer gain on the error of the estimated output to drive state estimations to the true state. 

For nonlinear systems, observers can be designed using various approaches, one of which is the \emph{Kazantzis-Kravaris/Luenberger (KKL) observer}, a generalization of the classical Luenberger approach for autonomous linear systems to nonlinear systems \cite{KKL}. 
The construction of a KKL observer involves transforming or lifting a nonlinear autonomous system into a linear time-invariant (LTI) system using an \emph{injective mapping} $T$, which satisfies a partial differential equation (PDE) condition. 
Then, the KKL observer is designed according to the system dynamics in the transformed coordinates. Unlike the Luenberger observer, the KKL observer does not use a direct gain matrix on the innovation quantity, but a pseudoinverse of the injective mapping therein must be found to recover the estimated system states. 
To guarantee the existence of such an injection, the backward distinguishability conditions and the admissible choice of the observer's LTI dynamics have been studied in theoretical works, e.g., \cite{Andrieu, Brivardis, Pachy}. 

\par While analytical constructions provide theoretical interpretability, the KKL observer remains challenging to implement because of the difficulty in numerically solving PDEs for the injective mapping and finding its pseudoinverse.  
Moreover, the computation for observer synthesis requires the knowledge of the mathematical model, which is often unknown or inaccurate. 
This limitation prompts many to pursue data-driven methods relying on tools, such as neural networks, physics-informed neural networks, and Lipschitz-bounded neural networks, to construct the KKL injective mapping \cite{NN, PINN, LBNN}. 
Non-neural approaches, including an online learning-based approach to learn the Chen-Fliess series expansion of the observer \cite{OnlineFliess}, and a kernel canonical correlation analysis approach \cite{KCCA}, have been proposed with the purpose of avoiding nonconvex training procedures.  

\par The Koopman operator, defined for the state-space dynamics, acts on a function space to advance them with the flow of the system \cite{Koopman}. In particular, we focus on the Koopman operator’s action on the span of its eigenfunctions. These eigenfunctions provide a linearizing coordinate transform of the original state dynamics, i.e., the evolution of the Koopman eigenfunctions is governed by an exponential factor of the eigenfunction values at an initial time point. 
\emph{If Koopman eigenfunctions can be identified from data}, then information about the system’s behavior can be obtained without knowledge of the first-principles mathematical model, unlike in analytical methods. The Koopman operator is hence a useful tool for data-driven KKL observer synthesis, given that it expresses nonlinear systems as infinite-dimensional linear systems, aligning with the purpose of the KKL injective map. 
Such an idea was preliminarily discussed in \cite{Surana}, where the system is assumed to be linearizable by a finite set of Koopman eigenfunctions, who span exactly contains all state components and output functions. Theoretically, such assumptions are unwarranted and hence the approach was only formally valid. In a recent work \cite{Ye}, using extended dynamic mode decomposition for Koopman operator estimation, the probabilistic error bounds are incorporated in a robust linear observer synthesis formulation. 

In this work, we propose an observer design that utilizes the Koopman eigenfunctions to explicitly build the injective map of the KKL observer for limit cycle systems from data. As proved in Mezi{\'{c}} \cite{Mezic}, a limit cycle system, where initial states in a domain of attraction converge to a closed planar curve, has Koopman eigenfunctions that correspond to real and imaginary eigenvalues as a result of its periodic and contracting nature. These Koopman eigenfunctions \cite{Mezic} form a rich basis of state-dependent functions. 
\begin{itemize}
\item First, we generate data from a limit cycle system to estimate the Koopman eigenfunctions, each associated with a real or imaginary eigenvalue, as a linear combination of a set of basis functions, using a least-squares regression problem solved as an eigenvalue minimization problem.
\item Second, the KKL observer, taking inspiration from \cite{Pachy}, is set to be multiple linear filters of the output. Each component of the injection $T$, satisfying this contracting filter for a chosen convergence speed parameter, is hence as a linear combination of the complex Koopman eigenfunctions using a finite difference approximation for the injective mapping time derivative. The linear combination coefficients are then determined using least-squares regression. 
\item Third, the left pseudoinverse of the injection $T$, mapping the observer state to the estimated system state, is estimated using kernel ridge regression (KRR) \cite{KRR}, which still forms a convex optimization problem.
\end{itemize}

Many advantages are offered by this method, with one being that a data-driven estimation of the injective map is provided. Unlike analytical methods, the proposed method eliminates the need for a mathematical model. Both tools utilized in this method, least squares regression and kernel ridge regression, require much less computation time than neural networks and avoid nonconvex or stochastic training procedures. While omitted in the present paper, these regression routines enjoy theoretical guarantees on their generalized performance, as is well known in statistical learning theory, which can be used to establish the state observer's probabilistic error bound. 

\section{PRELIMINARIES}
\subsection{KKL Observer Theory}
Consider a nonlinear autonomous system
\begin{equation}\label{system}
\dot{x} = f(x), \enspace y = h(x) \end{equation}
defined on a bounded open region $\Omega \subset \mathbb{R}^{n_x}$ where $x \in \Omega$ is the system state, and $y \in \mathbb{R}^{n_y}$ is the measured output. The goal of the KKL observer is to transform (\ref{system}) into the form 
\begin{equation}\label{zdot}
\dot{z} = Az + By
\end{equation}
such that $A \in \mathbb{R}^{n_z \times n_z}$ is a Hurwitz matrix, $B \in \mathbb{R}^{n_z \times n_y}$ is a matrix, and $z\in\mathbb{R}^{n_z}$ is the transformed coordinates obtained from the injective mapping $T: \Omega \to \mathbb{R}^{n_z}$ of the original state $x$ satisfying the PDE
\begin{equation}\label{KKL}
\frac{\partial T}{\partial x}f(x) = AT(x) + Bh(x).
\end{equation}
This PDE is a consequence of replacing $z$ in (\ref{zdot}) with the equivalent $T(x)$. Notice that the left-hand side of (\ref{KKL}) is the expansion of the injective mapping time derivative and also equal to $\mathcal{L}_fT$, the Lie derivative of $T$ along $f$. As this is a Sylvester-like PDE, it would be difficult to solve for an explicit expression of $T(x)$.

The KKL observer is thus constructed as a copy of (\ref{zdot}), 
\begin{equation}\label{KKL-observer}
\dot{\hat{z}} = A\hat{z} + By, \enspace \hat{x} = T^{\dagger}(z),
\end{equation}
and the estimated states recovered by applying $T^{\dagger}:\mathbb{R}^{n_z} \to \mathbb{R}^{n_x}$, the left pseudoinverse of the injective mapping, on the transformed coordinate $z$. To guarantee that $T$ is injective, (\ref{system}) must satisfy a backward distinguishability condition.

\begin{definition}\label{def:backdis} 
On an open set $\Omega$, the system (\ref{system}) is said to be backward distinguishable, if for any pair of solutions $(x_a, x_b)$ initialized in $\Omega$, there exists $t<0$ such that $(x_a, x_b)$ is defined in $\Omega$ on $[t, 0]$ and $h(x_a(t)) \neq h(x_b(t))$.
\end{definition}
By satisfying backward distinguishability on $\Omega$, it can be said that the signal of $y=h(x)$ holds enough information to identify a unique initial condition for each solution. 

\subsection{Koopman Operator for Dynamical Systems}
The Koopman operator, $K: \mathcal{G}(\Omega) \to \mathcal{G}(\Omega)$, defined for a discrete-time dynamical system $x_{k+1}=F(x_k)$ on a space of functions defined on $\Omega$, $\mathcal{G}(\Omega)$, is defined as such a mapping:
\begin{equation}\label{eq:Koopman}
Kg = g\circ F, \enskip \forall g\in \mathcal{G}(\Omega).
\end{equation} 
For example, if $F$ is continuous, then we can define $\mathcal{G}(\Omega)$ as the set of continuous functions on $\Omega$. In particular, we are interested in the action of the Koopman operator on the span of its eigenfunctions. The Koopman eigenfunction, $\varphi$, associated with eigenvalue $\rho \in \mathbb{C}$ is defined as a function that satisfies
\begin{equation}\label{eq:Koopeig}
\varphi(x_{k+1}) = K\varphi(x_k) = \rho\varphi(x_k) , \enskip k \in \mathbb{N}.
\end{equation}
The equation above shows that the Koopman operator scales its eigenfunction by its eigenvalue to advance it forward in time. The eigenfunction, at any two discrete time instants, $k$ and $k + \Delta k$, has the relation $\varphi(x_{k+\Delta k}) = \rho^{\Delta k}\varphi(x_k)$.

So far the Koopman operator has been defined for discrete-time systems. When working with continuous-time systems like in (\ref{system}),  we need a continuous counterpart of the Koopman operator. The action of the continuous counterpart of the Koopman operator, the Lie operator $L: \mathcal{G}(\Omega) \to \mathcal{G}(\Omega)$, on any $g \in \mathcal{D}(L) \subset \mathcal{G}( \Omega)$ is defined as
\begin{equation}\label{kooptolie}
Lg = \lim_{t \rightarrow 0}\frac{K^tg - g}{t} =  \lim_{t \rightarrow 0}\frac{g\circ F^t - g}{t}.
\end{equation}
The right-hand side of (\ref{kooptolie}) is known as the Lie derivative $\mathcal{L}_fg =\nabla g \cdot f $.  
If $L$ has its eigenfunctions $\varphi$ and corresponding eigenvalues $\mu$, i.e., $L\varphi = \mu\varphi$, then $K^t\varphi = e^{\mu t}\varphi$ for $t\geq0$. Thus, for a given sampling interval $\Delta t$, the corresponding discrete-time system has an eigenvalue $e^{\mu\Delta t}$.
\begin{remark}\label{remark:eig_mult}
The product of two Koopman eigenfunctions $\varphi_1(x)$ and $\varphi_2(x)$ corresponding to the eigenvalues $e^{\mu_1t}$ and $e^{\mu_2t}$ is also a Koopman eigenfunction $\varphi_1(x)\varphi_2(x)$ with the eigenvalue $e^{(\mu_1+\mu_2)t}$.  
Thus, if a system has a Koopman eigenfunction, then it has an infinite set of Koopman eigenfunctions, assuming that the function space $\mathcal{G}(\Omega)$ is closed under multiplication. Consequently, the Koopman operator may express a nonlinear system as an infinite-dimensional linear system, when such an infinite set of eigenfunctions form a ``basis" of the function space. 
\end{remark}

\subsection{Koopman Eigenfunctions of Limit Cycle Systems}
Let us consider (\ref{system}) as a two-dimensional stable limit cycle system with a domain of attraction $\Omega$. Then, (\ref{system}) inside $\Omega$ can be rewritten in the form:
\begin{equation}\label{ys_sys}
\dot{y} = A(s)y, \enskip \dot{s} = 1
\end{equation}
for $y\in \mathbb{R}, s\in \mathbb{S}^1 = \mathbb{R}/(2\pi \mathbb{Z})$ (i.e., $s$ is on a unit circle), and a $2\pi$-periodic function $A(s)$. The transformed variables $y$ and $s$ can be seen as representing the radial and angular movement of (\ref{system}) around the limit cycle, respectively. As the limit cycle is periodic, the system must be periodic in $s$, which is followed by $\dot{y}$ as it depends on the periodic function $A(s)$.
As proved in \cite{Mezic}, any function analytic in $y$ and $L^2$ (square-integrable) in $s$ can be expanded into the Koopman eigenfunctions of (\ref{ys_sys}). 

\begin{theorem}[Mezić \cite{Mezic}]
If a function $h(y, s)$ is analytic in $y$ and $L^2$ in $s$, then it can be expanded into the Koopman eigenfunctions of the system  (\ref{ys_sys}) in the following manner:
\begin{equation}\label{eig_expansion}
h(y, s)  = \sum_{m = 0 }^{\infty}\sum_{n = -\infty}^{\infty}c_{m n}\varphi(y, s; m, n),\enskip c_{mn}\in\mathbb{C},
\end{equation}
where
$$\varphi(y, s; m, n) =y^m\exp\left[-m\int^s_0{(A(\bar{s})-A^\ast)d{\bar{s}}}\right]e^{ins}, $$
$$ A^* = \frac{1}{2\pi}\int_0^{2\pi}A(z)dz. $$
Here, $\varphi(y, s; m, n)$ are Koopman eigenfunctions, each associated with the eigenvalue $e^{(mA^* + in)t}$ and  $c_{mn}$ are constant coefficients. $A^*$ is the average of $A(s)$, a real-valued function as $y\in\mathbb{R}$, over one period. 
\end{theorem}

\par To show that  $\varphi(y, s; m, n)$ and  $e^{(mA^* + in)t}$ are indeed Koopman eigenfunctions and associated eigenvalues, recall that a Koopman eigenfunction of (\ref{ys_sys}) must satisfy
\begin{equation}\label{ys_eigfunc}
\phi(y(t), s(t)) = \mu(t)\phi(y_0, s_0)
\end{equation}
for some $\mu(t) \in \mathbb{C}$ and
$$ s(t) = t + s_0, \enskip y(t) = y_0 \exp\left(\int_0^tA(\bar{t}+s_0)d\bar{t}\right).$$
The readers can verify that for $\mu(t) = e^{mA^*}$, $\phi(y, s) = y^m \exp(-m\int^s_0{(A(\bar{s})-A^\ast)d{\bar{s}}})$ satisfies (\ref{ys_eigfunc}). Furthermore, $\phi(s) = e^{ins}$ is also Koopman eigenfunction associated with the eigenvalue $\mu(t) = e^{int}$ since $K^te^{ins} = e^{in(s+t)}=e^{int}e^{ins}.$
Because the product of two eigenfunctions is also a Koopman eigenfunction (see Remark \ref{remark:eig_mult}), 
$$\varphi(y, s; m, n) = y^m \exp\left[-m\int^s_0{(A(\bar{s})-A^\ast)d{\bar{s}}}\right]e^{ins} $$
are also eigenfunctions associated with the eigenvalues $e^{(mA^* + in)t}$. 

\par Now that the existence and form of the Koopman eigenfunctions has been established, we next explain that $h(y, s)$ can be written as the infinite sum presented in (\ref{eig_expansion}). First, the analyticity of $h(y, s)$ in $y$ allows us to expand $h(y, s)$ as a Taylor series in $y$:
$$ h(y, s) = \sum_{m=0}^\infty a_m(s)y^m$$
such that $a_m(s)$ is $L^2$ and a $2\pi$-periodic function of $s$. The periodic function can then be expanded as Fourier series in $s$ as follows:
$$a_m(s) = \exp\left[-m\int^s_0{(A(\bar{s})-A^\ast)d{\bar{s}}}\right]\bar{a}_m(s),$$
where 
$$\bar{a}_m(s) = \sum_{n = -\infty}^{\infty}a_{mn}e^{ins},$$
since $ \exp(-m\int^s_0{(A(\bar{s})-A^\ast)d{\bar{s}}})$ is a positive and bounded function. The function being positive allows $\bar{a}_m(s)$ to be $L^2$, and thus expressed as a Fourier series, and bounded.

\section{Constructing the KKL Injective Map}
Consider (\ref{system}) for $n_y = 1$. We make the following assumptions. 
\begin{assumption}\label{assumption:back_dis}
The nonlinear system is backward distinguishable on $\Omega$ and strongly differentially observable, i.e., there exists an $n_z\in\mathbb{N}$ such that
the mapping $H: \mathbb{R}^{n_x} \rightarrow \mathbb{R}^{n_z}$ defined by 
$$ H(x) = \begin{bmatrix} h(x) \\ \mathcal{L}_fh(x) \\ \vdots \\ \mathcal{L}_f^{n_z - 1}h(x) \end{bmatrix} $$
is $k$-Lipchitz injective on $\Omega$ for some $k>0$.
\end{assumption}

\begin{assumption}\label{assumption:compact}
Let $\Omega_0 \in \mathbb{R}^{n_x}$ be the set of initial states $x_0$. The domain $\Omega \in \mathbb{R}^{n_x}$ is a compact set such that all solutions $x(t)$ with $x_0 \in \Omega_0$ remain in $\Omega$ for $t \geq 0$.
\end{assumption}

\begin{assumption}\label{assumption:output}
The output function $h(x) = h(y, s)$ is analytic in $y$ and $L^2$ in $s$.
\end{assumption}

\subsection{Construction of $T$}
In the recent work of Pachy et al. \cite{Pachy}, it was proposed to construct the observer dynamics as nonlinear contracting filters of the output in parallel, i.e.,
\begin{equation}\label{eq:pachykkl}
\dot{z} = k\boldsymbol{\sigma}(z, y), \enskip k >0,
\end{equation}
in which 
$$\boldsymbol{\sigma}(z, y) = (\lambda_1\sigma(z_{1}, y), \cdots, \lambda_{n_z}\sigma(z_{n_z}, y)).$$ 
Here $\{\lambda_i\}_{i = 1}^{n_z}$ are positive scalars that specify the observer convergence speed and $\sigma: \mathbb{R}\times\mathbb{R}\rightarrow\mathbb{R}$ is a scalar-valued, exponentially contracting function satisfying:
\begin{equation}\label{eq:pachy_crit}
0 < \gamma \leq \left| \frac{\partial\sigma}{\partial y}(\zeta, y)\right|, \enspace
- \beta \leq \frac{\partial \sigma}{\partial \zeta} (\zeta, y) \leq -\alpha < 0,
\end{equation}
for all $\zeta, y \in \mathbb{R}$. It is proved in \cite{Pachy} that under Assumption \ref{assumption:back_dis} and Assumption \ref{assumption:compact}, there exists an injective mapping $z=T(x)$, defined on $\Omega$, that gives the KKL observer in the form of (\ref{eq:pachykkl}).

Although these criterias were intended for nonlinear $\sigma$ functions in \cite{Pachy}, it is clear that a linear $\sigma$ function suffices to satisfy (\ref{eq:pachy_crit}). Hence, (\ref{eq:pachykkl}) provides a nonlinear generalization of (\ref{zdot}). Consider, for example,
$$\sigma(z_j, y) = -c_1z_j + c_2y, \enskip c_1 > 0, c_2 \neq 0.$$
If we let $k =1$, $c_1 = 1$, $c_2 = 1$, and rescale each $z_j$ by $\lambda_j$, then linear dynamics in the form (\ref{zdot}) can be obtained, where 
\begin{equation}\label{eq:pachy_lin}
        A = \begin{bmatrix}-\lambda_1&0&\cdots&0\\0&-\lambda_2&\ddots&\vdots\\\vdots&\ddots&\ddots&0\\0&\cdots&0&-\lambda_m\end{bmatrix}, \enskip B = \begin{bmatrix}1\\1\\\vdots\\1\end{bmatrix},
    \end{equation}
i.e., $\dot{z}_j = -\lambda_j z_j + y$ for all $j \in \{1, \cdots, n_z\}$. 
With the observer design outlined, we now provide the construction for the injective mapping $T$ based on Assumption \ref{assumption:output}. 
\begin{theorem}\label{thm:2}
Assume $h(x) = h(y, s)$ is analytic in $y$ and $L^2$ in $s$. Then, $h$ can be written as a linear combination of Koopman eigenfunctions, i.e., $h = \sum_i \eta_i \varphi_{\mu_i},$ and hence if $\lambda_j \neq -\mu_i$ for all $j\in\{1, \cdots, n_z\}$ are satisfied for the selected parameters $\lambda_1, \cdots, \lambda_{n_z}$, then the components of the injective mapping are also linear combinations of Koopman eigenfunctions.
\end{theorem}
\begin{proof}
Given $A$ and $B$ defined in (\ref{eq:pachy_lin}), each component of (\ref{KKL}) becomes 
\begin{equation}\label{eq:component_KKL} \frac{\partial T_j}{\partial x} \cdot f =  -\lambda_jT_j + h. \end{equation}
Since the left-hand side is equivalent to the Lie derivative $\mathcal{L}_fT_j$ and $h = \sum_i \eta_i \varphi_{\mu_i}$, the PDE can be rewritten as 
$$ (\mathcal{L}_f + \lambda_j)T_j = \sum_i \eta_i \varphi_{\mu_i}.$$
Note that all Koopman eigenfunctions, $\varphi_{\mu_i}$, are eigenfunctions of $(\mathcal{L}_f + \lambda_j)$. For an ODE, the particular solution is a multiple of its eigenfunction if the right-hand side function is also a multiple of the eigenfunction. 
Based on this idea, we can set $T_j = \sum_iC_i\varphi_{\mu_i}$, and the PDE now becomes
$$ (\mathcal{L}_f + \lambda_j)\sum_i C_i \varphi_{\mu_i}= \sum_i \eta_i \varphi_{\mu_i}.$$
Since $\mathcal{L}_f\varphi_{\mu_i} = \mu_i \varphi_{\mu_i},$
$$\sum_i C_i (\mu_i + \lambda_j) \varphi_{\mu_i}= \sum_i \eta_i \varphi_{\mu_i}.$$
Thus, $C_i = \frac{\eta_i}{\mu_i + \lambda_j}$, namely $T_j = \sum_i\frac{\eta_i}{\mu_i+\lambda_j}\varphi_{\mu_i}$.
\end{proof} 

\subsection{Estimation of Koopman Eigenfunctions}
Without a first-principles model, we turn to Koopman eigenfunctions to obtain information about the system (\ref{system}) and find a linearizing coordinate transformation. Below, we describe Algorithm \ref{alg:eig}, which uses a linear regression routine to estimate a Koopman eigenfunction associated with a specified eigenvalue $\mu$. 

\par Let us take two arrays of data points, collected from multiple trajectories of the dynamical system, $X = [\mathbf{x}_1, \cdots, \mathbf{x}_d]^{\top}$ and $X^+ = [\mathbf{x}_1^+, \cdots, \mathbf{x}_d^+]^{\top}$ where for each $i \in \{1, \cdots, d\}$, $\mathbf{x}_i = x(t_i), \enskip \mathbf{x}_i^+ = x(t_i + \Delta t)$ for a fixed sampling interval $\Delta t > 0$. From (\ref{eq:Koopeig}) and the relation between the continuous-and discrete-time eigenvalues, $\rho = e^{\mu\Delta t}$, we obtain 
\begin{equation}\label{eq:Koopeig_time}
\varphi(\mathbf{x}_i^+) = K^{\Delta t}\varphi(\mathbf{x}_i) = e^{\mu\Delta t}\varphi(\mathbf{x}_i). 
\end{equation}
Therefore, to estimate the Koopman eigenfunctions of (\ref{system}), we must find functions which satisfies (\ref{eq:Koopeig_time}). 

\par Suppose that $\varphi \in \operatorname{span}\{g_i\}_{i = 1}^{n_G}$ where $\{g_i\}_{i=1}^{n_G}$ is a set of basis functions, e.g., polynomial functions up to a user-specified degree. 
Let $G = [g_1, g_2, \cdots, g_{n_G}]^{\top}$, a $n_G$-dimensional vector-valued function. Then, $\varphi(x) = G^{\top}(x)\beta$ for some $\beta\in \mathbb{C}^{n_G}$. The problem now becomes identifying such a vector $\beta$ that satisfies $G^{\top}(\mathbf{x}_i^+) \beta= e^{\mu\Delta t}G^{\top}(\mathbf{x}_i)\beta,$
i.e.,
\begin{equation}
\gamma_i^\top \beta= 0, \enspace i\in\{1, 2, \cdots, d\}, 
\end{equation}
where we denote $\gamma_i = G(\mathbf{x}_i^+) - e^{\mu\Delta t}G(\mathbf{x}_i)$. While generally, $\varphi \notin \operatorname{span}\{g_i\}^{n_G}_{i=1}$, for approximation, we can construct the minimization problem
$$ \min_\beta \sum_{i =1}^{d} |\gamma_i^{\top}\beta|^2 = \beta^* \Gamma \beta, \enspace \mathrm{s.t.}\, \|\beta\| =1$$ 
where the norm constraint ensures $\beta$ does not become trivially zero and $\Gamma = \sum_{i = 1}^{d} \bar{\gamma}_i \gamma_i^{\top}$. (For a complex vector $v$, the notation for its transpose is $v^{\top}$, conjugate is $\bar{v}$, and conjugate transpose is $v^*$.) 

\par Then $\Gamma$ is an Hermitian matrix which can be decomposed into $\Gamma = U\Lambda U^*$ with a unitary matrix $U$ and diagonal matrix $\Lambda$. Each column vector of $U$, $U_i$, is an eigenvector of $\Gamma$ with the associated eigenvalue $\Lambda_{ii} \in \mathbb{R}$. 
Thus, if we let $\beta= U_i$, the cost becomes $\Lambda_{ii}(U_i^*U_i)^2 = \Lambda_{ii}$, which can be minimized by taking $\Lambda_{ii} = \lambda_{\min}(\Gamma)$. Notice that the criteria $\|\beta\| =1$ remains satisfied as the eigenvectors that $U$ is comprised of are orthonormal.

\begin{algorithm}[!t]
\caption{Regression for Koopman eigenfunctions}\label{alg:eig}
\begin{algorithmic}[1]
\State Input: $G = [g_1, \cdots g_n]$, $\Delta t$, $\mu$, $X$, $X^+$
\State Compute $\gamma_i = G(\mathbf{x}_i^+) - e^{\mu\Delta t}G(\mathbf{x}_i)$ for $i=1,\dots,d$, and subsequently $\Gamma = \sum_{i = 1}^{d} \bar{\gamma}_i \gamma_i^{\top}$.
\State Obtain $(U, \Lambda)$ from the eigenvalue decomposition of $\Gamma$.
\State Determine $p = \arg \min_i \Lambda_{ii}$ and set $\beta = U_p$.
\State Output: $ \varphi(\cdot) = G^{\top}(\cdot)\beta$
\end{algorithmic}
\end{algorithm}

\subsection{Determination of $T$ and $T^{\dagger}$}
In Theorem \ref{thm:2}, it was proven that each component of the injection, $T_j$ ($j=1,\dots,n_z$), is a linear combination of the Koopman eigenfunctions $\varphi_{\mu_i}$ for $T_j$ that satisfies the PDE (\ref{eq:component_KKL}). Specifically, the eigenvalues to be used are of the form $m\mu_{\text{real}} + n\mu_{\text{imag}}$, where $\mu_{\text{real}}$ is negative and real and $\mu_{\text{imag}}$ is imaginary. 
Suppose that $\mu_{\text{real}}$ and $\mu_{\text{imag}}$ are given. To find the coefficients in $T_j$ as linear combinations, we can take a finite difference approximation of the left-hand side in (\ref{eq:component_KKL}) to obtain 
\begin{equation}
\frac{T_j(x(t+\Delta t)) - (1+\lambda_j \Delta t)T_j(x(t))}{\Delta t} = h(x(t)).
\end{equation}
Now, let us approximate $T_j$ as a linear combination of Koopman eigenfunctions $T_j = b_j^{\top}\boldsymbol{\varphi}$. Then, given the data arrays $X$ and $X^+$, the minimization problem becomes
\begin{equation}\label{eq:min_T}
\min_{b_j}\sum_{i=1}^{d}\bigg|\bigg| b_j^{\top}\left[ \frac{\boldsymbol{\varphi}(\mathbf{x}_i^+) - (1+\lambda_j \Delta t)\boldsymbol{\varphi}(\mathbf{x}_i)}{\Delta t}\right] - h(\mathbf{x}_i) \bigg|\bigg|^2,
\end{equation}
which is a least-squares regression problem. The regression for KKL injection mapping $T$ is given in Algorithm \ref{alg:T}. 

\begin{algorithm}[!t]
\caption{Regression for $T$}\label{alg:T}
\begin{algorithmic}[1]
\State Input: $G = [g_1, \cdots g_n]$, $\Delta t$, $\mu_{\text{real}}$,  $\mu_{\text{imag}}$, $X$, $X^+$, $M$, $N$
\State Estimate the eigenfunctions $\{\varphi_{\text{real}, m}\}_{m=0}^{M}$ and $\{\varphi_{\text{imag}, n}\}_{n=-N}^{N}$ corresponding to the eigenvalues $\mu_{\text{real}}^m$ for $m \in \{0, 1, 2, \cdots, M\}$ and $\mu_{\text{imag}}^n$ for $n \in \{-N, -N+1, \cdots, 0, 1, \cdots, N\}$ using Algorithm 1. 
\State Form the set of complex Koopman eigenfunctions,  $\{\varphi_{\text{real}, m}\varphi_{\text{imag}, n}: 0\leq m \leq M,  |n| \leq N\}$ by taking the pairwise products of real and imaginary Koopman eigenfunctions.
\State Let $\boldsymbol{\varphi}$ be the vector of complex Koopman eigenfunctions. Compute $v_i = \left[\boldsymbol{\varphi}(\mathbf{x}_i^+) - (1+\lambda_j \Delta t)\boldsymbol{\varphi}(\mathbf{x}_i)\right]/ \Delta t$.
\State Determine $b_j^* =\arg \min_{b_j}\sum_{i=1}^{d}\| b_j^{\top}v_i - h(\mathbf{x}_i) \|^2$ for $j = 1, \cdots, n_z$.
\State Output: $T_j = b^{*\top}_j\boldsymbol{\varphi}$, \enskip $T = [T_1, \cdots, T_{n_z}]$
\end{algorithmic}
\end{algorithm}

To find the inverse of the injective mapping, $T^{\dagger}$, here we use the technique of kernel ridge regression (KRR). Once the injection $T$ has been estimated, then for data points $\{\mathbf{x}_i\}_{i=1}^d$, evaluating them using the estimated $T$ gives $\{\mathbf{z}_i=T(x_i)\}_{i=1}^d$. Hence, the estimation of $T^\dagger$ is a nonlinear regression problem with $\{\mathbf{z}_i\}_{i=1}^d$ being the input data and $\{\mathbf{x}_i\}_{i=1}^d$ being the output labels. 
The KRR approach is a generic one for nonlinear regression, which bypasses the explicit choice of a set of basis functions on the $\mathbf{z}$-data; instead, we use a kernel function kernel $Q(\cdot, \cdot)$, which is a continuous, real-valued bivariate function on $\mathbb{R}^{n_z}$ such that the kernel matrix, $\mathbf{Q} \in \mathbb{R}^{d \times d}$ where $\mathbf{Q}_{ij} = Q(\mathbf{z}_i, \mathbf{z}_j)$, is guaranteed to be positive definite for any possible choice of $\mathbf{z}_1, \dots, \mathbf{z}_d$ and $d\in \mathbb{N}$. 

\par The KRR involves solving for 
$$ w^* = \arg \min_{w} \frac{1}{d} \sum_{i=1}^{d}\|\mathbf{x}_i - w^{\top}\Psi(\mathbf{z}_i)\|^2 + \xi\| w\|^2 $$
where the regularization parameter, $\xi$, prevents overfitting by shrinking coefficients. $\Psi = [\psi_1, \psi_2, \cdots, \psi_{n_\Psi}]$ is formally a vector of basis functions forming a linear combination that approximates $T^{\dagger}$ and satisfying $Q(\mathbf{z}_i, \mathbf{z}_j) = \Psi^\top(\mathbf{z}_i)\Psi(\mathbf{z}_j)$. 
By duality principle, $w^{\top}\Psi$ can be expressed as $\sum_i\alpha_i^* Q(z_i, \cdot)$, which converts KRR intoa new minimization problem on $\alpha$: 
\begin{equation}\label{eq:opt_alpha}
\alpha^* = \arg \min_{\alpha} \frac{1}{d} \sum_{i=1}^{d}\|\mathbf{x}_i - \mathbf{Q}_i\alpha\|^2 + \xi \operatorname{Tr}(\alpha\mathbf{Q}\alpha^{\top}). 
\end{equation}
Here $\alpha \in \mathbb{R}^{d \times n_x}$ and $\mathbf{Q}_i$ is the $i$-th row of $\mathbf{Q}$. The readers can verify that $\alpha^* = (\mathbf{Q} + d\xi I)^{-1}X$ and hence $T^{\dagger}(z) =  \sum_{i = 1}^d\alpha_i^* Q(z_i, z)$, where $\alpha_i$ is the $i$-th row of $\alpha$. 

\begin{algorithm}[!t]
\caption{Kernel ridge regression for $T^{\dagger}$}\label{alg:KRR}
\begin{algorithmic}[1]
\State Input: $\xi$, $X$, $Q$ (kernel function), $T$
\State Create an array of observer states $\mathbf{z} = T(\mathbf{x})$.
\State Compute $\mathbf{Q}_{ij} = Q(\mathbf{z}_i, \mathbf{z}_j)$ to identify $\alpha^*$ from (\ref{eq:opt_alpha}).
\State Output: $T^{\dagger} = \sum_{i = 1}^d\alpha_i^* Q(\mathbf{z}_i, \cdot)$
\end{algorithmic}
\end{algorithm}

\section{Numerical Results}
Consider the Brusselator system,
\begin{equation}\label{eq:brusselator}
\dot{x}_1 = a + x_1^2x_2 - (b+1)x_1, \enskip \dot{x}_2 =bx_1 - x_1^2x_2,
\end{equation}
where $x^* = (a, b/a)$ is the equilibrium point. Let $a = 1, b = 3$. For simulation, $100$ trajectories are generated, each with a time duration of $3.0$ and a sampling interval of $\Delta t = 0.1$. The initial states for the trajectories are sampled from a normal distribution $\mathcal{N}(x^*, 0.75^2I_2)$, and filtered with the following criteria to avoid data close to the axes or faraway:
\begin{equation*}
x_1 \geq 0.2, \enskip x_2 \geq 0.1, \enskip \|x - x^*\| \geq 0.5, \enskip x_1 + x_2 \leq 7.
\end{equation*}
For each trajectory array, we let $X_i$ be the $i$-th trajectory with the final time point removed and $X_i^+$ be the $i$-th trajectory with the first time point removed. The arrays $(X_i)_{i = 1}^{l}$ are vertically concatenated to form $X$, and $X^+$ is obtained analogously.
 
Next, the basis functions are chosen to be multivariate polynomials of $x_1$ and $x_2$ with a maximum total degree of $5$ to estimate the Koopman eigenfunctions corresponding to $\{\mu_{\text{real}}^m\}_{m=0}^{M}$ and $\{\mu_{\text{imag}}^n\}_{n = -N}^{N}$. To choose $\mu_{\text{real}}$ and $\mu_{\text{imag}}$, we must identify some features of the Brusselator system. Since imaginary eigenvalues represent the oscillation frequency, we let $\mu_{\text{imag}} = \frac{2\pi}{P}i$, where $P$ is the period of the limit cycle. For a Brusselator (\ref{eq:brusselator}), $P \approx 7.16.$ For the real eigenvalue, the rate constant of (\ref{eq:brusselator}), set to $1$, is identified, and hence we let $\mu_{\text{real}} = -1$. For the remaining inputs of Algorithm \ref{alg:T}, we choose $M=7$ and $N=7$ to be sufficient for accurate estimations of the injective mapping $T$, verified by the RMSE of the least-squares regression. 

\par It follows that the inverse mapping, $T^{\dagger}$ is estimated with KRR (Algorithm \ref{alg:KRR}) using the Laplace kernel:
\begin{equation}
Q(\mathbf{z}_i, \mathbf{z}_j) = \exp\left(-\|\mathbf{z}_i -\mathbf{z}_j\|/l\right),
\end{equation}
where $l = 2$ is empirically tuned. The Laplace kernel was empirically found to perform better than the commonly used Gaussian kernel, possibly due to the fact that the reproducing kernel Hilbert space generated by the Laplace kernel, which is not a universal kernel, is smaller, which has an advantage in reducing overfitting. 
To optimize the KRR estimation accuracy, we sample $1000$ datapoints from a normal distribution $\mathcal{N}(x^*, 1.15^2I_2)$, with the filtering criteria $x_1>0.2$ and $x_2>0.1$, and take $\xi$ to be close to $0$.

\begin{figure}[!t]
\centering
\includegraphics[width = \columnwidth]{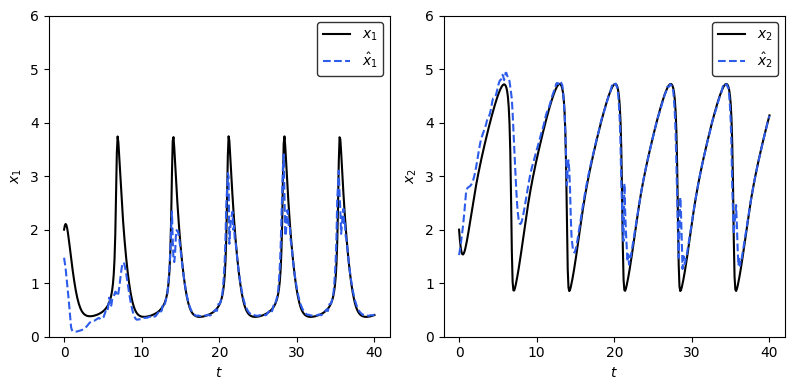}
\caption{Trajectories of states and observed states in the Brusselator.}
\label{fig:obs_result}
\end{figure}
\begin{figure}[!t]
\centering
\includegraphics[width = \columnwidth]{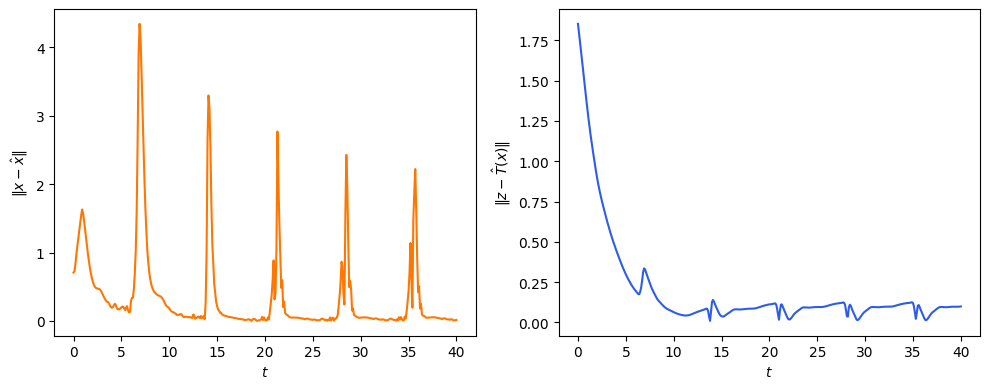}
\caption{$l_2$-norm error of \textit{(left)} estimated system states and \textit{(right)} observer state estimation of true transformed coordinate values.}
\label{fig:error}
\end{figure}
\begin{figure}[!t]
\centering
\includegraphics[width = \columnwidth]{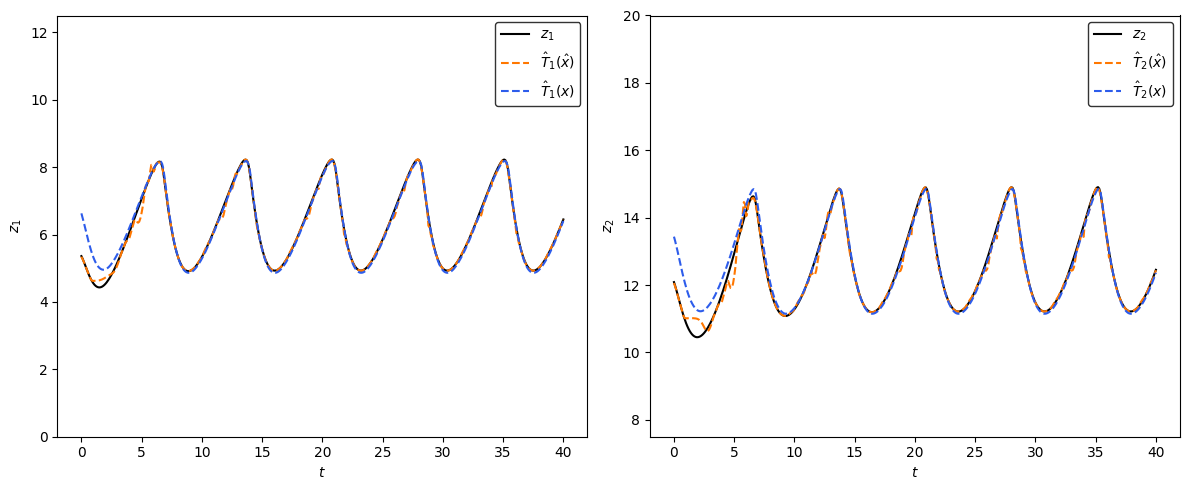}
\caption{Comparison of observer states, transformed true system states, and transformed estimated system states.}
\label{fig:obs_state}
\end{figure}

The observer is then constructed using (\ref{KKL-observer}) where $\lambda = (0.5, 0.25)$ ensuring that $\mu_i \neq -\lambda_j$. The trajectory to be estimated is initialized at $x_0 = (2, 2)$, and the observer is initialized at $T(\hat{x}_0)$ where $\hat{x}_0 = (1.5, 1.5)$. 
The resulting observer and system trajectories are compared in Figure \ref{fig:obs_result}, with the estimation errors, shown in Figure \ref{fig:error}, revealing that the estimated inverse mapping has most error at the trajectory peaks. 
The error is likely attributed to the KRR procedure; as naturally expected, the kernel approach contains a smoothing mechanism. This hypothesis is verified in Figure \ref{fig:obs_state}, since the transformed estimated states $\hat{T}(\hat{x})$ only slightly deviate from the observer states $z$.

The codes for the numerical experiments in this section are available at GitHub Repository: \url{https://github.com/angelani567/KoopmanObserver}.

\section{Conclusions}
In this work, we presented a data-driven Koopman operator-based approach to learn the KKL observer for planar limit cycle systems. The existence of Koopman eigenfunctions are guaranteed for such systems and form a rich basis of state-dependent functions, enabling the injective mapping to be estimated as a linear combination of these eigenfunctions. The inverse of the injective mapping is then constructed using KRR. 

The proposed approach is computationally efficient, relies solely on convex optimization problem, and requires no first-principles model. However, certain limitations remain as the observer design is restricted to planar systems. Due to Mezić \cite{Mezic}, possible extensions to higher-dimensional systems can be envisioned. Additionally, some prior knowledge is necessary to determine eigenvalues as imaginary eigenvalues are determined by the limit cycle period and real eigenvalues may require knowledge of the rate constants. Extending the current approach to non-autonomous systems and utilizing data-driven observers in output-feedback controllers will be carried out in future works.

\end{document}